\documentclass[aps,twocolumn,nofootinbib,superscriptaddress,floatfix]{revtex4}

\usepackage{color,amsmath,amssymb,graphicx,latexsym,subfigure}
\pdfoutput=1
\usepackage{multirow}
\usepackage{ulem}

\usepackage{float}
\usepackage{booktabs}
\usepackage{threeparttable}
\usepackage[colorlinks,linkcolor=green, anchorcolor=green,citecolor=green]{hyperref}
\usepackage{ulem,soul}
\usepackage{xcolor}

\newcommand{\be}{\begin{equation}}
\newcommand{\ee}{\end{equation}}
\newcommand{\ba}{\begin{eqnarray}}
\newcommand{\ea}{\end{eqnarray}}

\newcommand{\gsim}{\mathrel{\hbox{\rlap{\lower.55ex \hbox {$\sim$}}
			\kern-.3em \raise.4ex \hbox{$>$}}}}
\newcommand{\lsim}{\mathrel{\hbox{\rlap{\lower.55ex \hbox {$\sim$}}
			\kern-.3em \raise.4ex \hbox{$<$}}}}

\hypersetup{colorlinks=true,
	breaklinks=true,
	pdfstartview=Fit,
	linkcolor=blue,
	citecolor=blue,
	urlcolor=blue}

\begin{document}

\title{Probing the sound speed and clustering of dark energy}

\author{Yuhang Yang}
\affiliation{Department of Astronomy, School of Physical Sciences, University of Science and Technology of China, Hefei 230026, China}
\affiliation{CAS Key Laboratory for Research in Galaxies and Cosmology, School of Astronomy and Space Science, 
University of Science and Technology of China, Hefei 230026, China}

\author{Qingqing Wang}
\affiliation{Department of Astronomy, School of Physical Sciences, University of Science and Technology of China, Hefei 230026, China}
\affiliation{CAS Key Laboratory for Research in Galaxies and Cosmology, School of Astronomy and Space Science, 
University of Science and Technology of China, Hefei 230026, China}
\affiliation{Kavli IPMU (WPI), UTIAS, The University of Tokyo, Chiba 277-8583, Japan}

\author{Xin Ren}
\affiliation{Lanzhou Center for Theoretical Physics, Key Laboratory of Theoretical Physics of Gansu Province,
and Key Laboratory of Quantum Theory and Applications of MoE, Lanzhou University, Lanzhou, Gansu 730000, China}
\affiliation{Institute of Theoretical Physics \& Research Center of Gravitation, School of Physical Science and Technology, Lanzhou University, Lanzhou 730000, China}
\affiliation{Department of Astronomy, School of Physical Sciences, University of Science and Technology of China, Hefei 230026, China}
\affiliation{CAS Key Laboratory for Research in Galaxies and Cosmology, School of Astronomy and Space Science, 
University of Science and Technology of China, Hefei 230026, China}

\author{Emmanuel N. Saridakis}
\email{msaridak@noa.gr}
\affiliation{National Observatory of Athens, Lofos Nymfon 11852, Greece}
\affiliation{CAS Key Laboratory for Research in Galaxies and Cosmology, School of Astronomy and Space Science, 
University of Science and Technology of China, Hefei 230026, China}
\affiliation{Departamento de Matem\'{a}ticas, Universidad Cat\'{o}lica del Norte,  Casilla 1280, Chile}

\author{Yi-Fu Cai} 
\email{yifucai@ustc.edu.cn}
\affiliation{Department of Astronomy, School of Physical Sciences, University of Science and Technology of China, Hefei 230026, China}
\affiliation{CAS Key Laboratory for Research in Galaxies and Cosmology, School of Astronomy and Space Science, 
University of Science and Technology of China, Hefei 230026, China}

\begin{abstract}
Recent Dark Energy Spectroscopic Instrument (DESI) observations favor a dynamical dark energy component with a 
time-varying equation-of-state, potentially crossing the cosmological-constant boundary \(w=-1\), challenging the standard \(\Lambda\)CDM paradigm. In this paper we present the first joint observational constraints on the clustering properties of such dynamical dark energy, using both the Parameterized Post-Friedmann (PPF) framework and the effective field theory (EFT) of dark energy. Combining DESI DR2 baryon acoustic oscillations with Planck 2018 cosmic microwave background data and the Union3 supernova sample, we constrain the effective sound speed \(c_s^{2}\). For a time-varying equation-of-state, the degeneracy between \((1+w)\) and \(c_s^{2}\) is broken, yielding the first meaningful constraint \(\log_{10}c_{s}^{2}=-3.00^{+2.9}_{-0.99}\), while constant-\(w\) models remain unconstrained. A complementary EFT analysis gives 
consistent results, favoring \(c_s^{2}\sim 0.3\) or \(0.4\). Our findings 
demonstrate that current data are now sensitive to the perturbative properties of dynamical dark energy, opening a new observational window on the nature of cosmic acceleration.

\end{abstract}


\maketitle


\section{Itroduction}

The standard $\Lambda$CDM paradigm has been greatly challenged by the recent observations from the Dark Energy Spectroscopic Instrument (DESI) Data Release~2 (DR2) \cite{DESI:2025zgx} by favoring a dynamical dark energy component when analyzed within the $w_0w_a$ (Chevallier-Polarski-Linder) parameterization \cite{Chevallier:2000qy, Linder:2002et} $w(a)=w_0+w_a(1-a)$. Depending on the supernova sample employed, deviations from $\Lambda$CDM at the level of $2.8$--$4.2\sigma$ have been reported, indicating that cosmic acceleration may be driven by physics beyond a cosmological constant. 
A particularly intriguing outcome of the DESI analysis is the robust preference for a dynamical evolution of dark energy, in which the equation-of-state parameter crosses the cosmological-constant boundary $w=-1$ during cosmic history \cite{Feng:2004ad, Hu:2004kh}. 
This behavior persists across different parameterizations and is also recovered in non-parametric reconstructions 
\cite{DESI:2025fii, DESI:2025wyn, Yang:2025kgc}, underscoring its data-driven character. 
The resulting tension with $\Lambda$CDM has stimulated extensive investigation from both observational and theoretical perspectives 
\cite{Efstathiou:2024xcq, DES:2025sig, Giare:2024smz, Ye:2024ywg, Lu:2025gki, Pan:2025qwy, Khoury:2025txd, Li:2025vuh}.

At the perturbative level, however, the aforementioned dynamics suffers a well-known theoretical challenge. In simple single-field or perfect-fluid descriptions of dark energy within General Relativity, crossing $w=-1$ leads to divergences in the perturbation equations, a result commonly referred to as the no-go theorem for quintom dark energy \cite{Cai:2009zp, Vikman:2004dc}. 
This difficulty can be resolved by involving multiple degree-of-freedom constructions or modified-gravity frameworks, such as Horndeski theories, where a finite effective sound speed or modified dispersion relations render the perturbations well behaved \cite{Guo:2004fq, Zhang:2005eg}. 
From a phenomenological viewpoint, the perturbative sector of dark energy has gained renewed importance, driven both by theoretical consistency requirements and by observational tensions in the growth of cosmic structure, most notably the $S_8$ tension \cite{DiValentino:2020vvd}.

In this context, the effective sound speed $c_s^2$ plays a central role. 
For dynamical dark energy, a sufficiently small sound speed suppresses the Jeans scale and allows dark energy perturbations to cluster on cosmological scales, thereby modifying the evolution of matter perturbations and leaving potentially observable imprints on large-scale structure and cosmic microwave background (CMB) observables 
\cite{Zhao:2005vj, Takada:2006xs, Xia:2007km, Basse:2012wd, Batista:2021uhb, Kunz:2015oqa}. 
Assessing the clustering properties of dark energy is therefore essential for evaluating the viability of dynamical equation-of-state models and for 
quantifying their impact on late-time observables. 
Given that DESI DR2 favors dynamical, and in particular quintom-like, dark energy, a consistent treatment of perturbations becomes a necessary ingredient of the analysis. 
A practical implementation of dark energy perturbations across the $w=-1$ 
boundary is provided by the Parameterized Post-Friedmann (PPF) framework, which models dark energy as a non-interacting fluid while ensuring regular behavior at phantom divide \cite{Fang:2008sn, Hu:2008zd}. 
A complementary description is offered by the effective field theory (EFT) of dark energy, which encompasses a broad class of scalar-field and 
modified-gravity models and allows perturbations to be studied independently of the background evolution 
\cite{Gubitosi:2012hu, Bloomfield:2012ff, Creminelli:2008wc}. 
With suitable stability conditions imposed, both approaches provide a 
theoretically consistent treatment of dark energy perturbations 
\cite{DeFelice:2016ucp}. 

Previous analyses, including those by the \textit{Planck} Collaboration, found the dark energy sound speed to be essentially unconstrained by available data \cite{Planck:2015bue}. 
In this paper, we revisit this question in light of the latest DESI DR2 
observations, in combination with CMB and supernova data. 
We show that once dark energy is allowed to be dynamical, current observations become sensitive to its perturbative properties, enabling meaningful constraints on the sound speed and clustering behavior of dark energy for the first time. 
\\


\section{The PPF approach}
At the perturbative level, dark energy can be 
treated as an additional non-interacting fluid, alongside cold dark matter and radiation, with equation-of-state $w=p_{\rm de}/\rho_{\rm de}$. 
In general, dark energy perturbations are non-adiabatic and may carry entropy (isocurvature) fluctuations. 
In an arbitrary gauge, the relation between the pressure and density 
perturbations of dark energy takes the form
\begin{equation}
\delta p_{\rm de}
=
c_s^2\,\delta\rho_{\rm de}
+
3(1+w)\rho_{\rm de}\left(c_s^2-c_a^2\right)\frac{v_{\rm de}}{k_H},
\end{equation}
where $k_H\equiv k/(aH)$, $c_s^2=\delta p_{\rm de}^{\rm rest}/\delta\rho_{\rm 
de}^{\rm rest}$ is the effective sound speed in the dark energy rest frame, and 
$c_a^2=\dot p_{\rm de}/\dot\rho_{\rm de}$ is the adiabatic sound speed.

For parameterized dark energy models that allow the equation-of-state to cross the cosmological-constant boundary $w=-1$, the adiabatic sound speed $c_a^2$ 
diverges at the crossing, leading to instabilities in the perturbation 
equations \cite{Vikman:2004dc,Deffayet:2010qz,Chimento:2008ws,Cai:2009zp}. 
Rather than addressing this issue at the level of specific microphysical 
models, here we adopt a phenomenologically consistent treatment of perturbations 
based on the PPF framework 
\cite{Fang:2008sn,Hu:2008zd}. 
In this approach, dark energy perturbations are described by a single dynamical 
variable $\Gamma$, constructed to remain regular across the phantom divide, 
while preserving local energy-momentum conservation.

A key assumption of the PPF formalism is that dark energy becomes effectively 
smooth relative to matter below a transition scale defined by $c_s k_H = 1$. 
This prescription allows the perturbation equations to be solved consistently 
on all relevant scales, yielding well-behaved metric and gravitational 
potentials even in the presence of quintom-like evolution. 
As a result, the framework provides a controlled and observationally viable 
description of dark energy perturbations in dynamical scenarios. See the
Appendix.~A for details of the PPF approach.

dark energy perturbations modify the evolution of the gravitational potentials and can therefore leave observable imprints, most notably through the late-time integrated Sachs-Wolfe (ISW) effect in CMB. 
Including these perturbations leads to a suppression of the CMB temperature 
power spectrum and of the matter power spectrum on large angular scales (low multipoles), while leaving small-scale observables essentially unaffected 
\cite{Fang:2008sn,Zhao:2005vj,Takada:2006xs}. 
These large-scale signatures are precisely where current CMB observations 
retain sensitivity to dark energy clustering.

In the following, we consider both constant-$w$ and dynamical $w_0w_a$ 
parameterizations. 
Models with $c_s^2=1$ in PPF framework correspond to smooth dark energy and are denoted as 
$w$CDM or $w_0w_a$CDM, respectively, where dark energy clustering on sub-horizon scales is strongly suppressed. 
Allowing for an arbitrary sound speed within the PPF framework, we refer to the 
corresponding scenarios as $w$CDM+PPF and $w_0w_a$CDM+PPF.\\


\begin{figure*}[htbp]
    \centering
    \includegraphics[width=\textwidth]{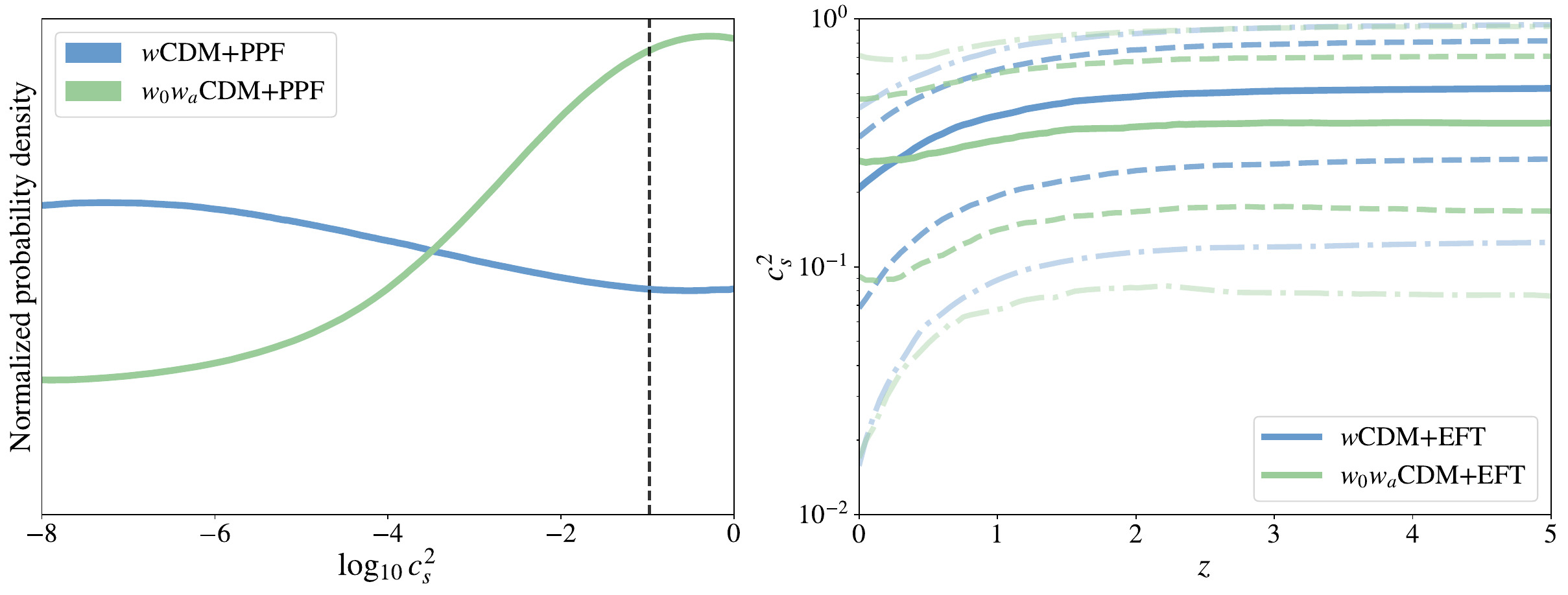}    
    \caption{{\it{Left: Normalized marginalized one–dimensional posteriors for 
    $\log_{10}c_s^2$ in 
    the $w$CDM+PPF and $w_0w_a$CDM+PPF models, obtained from the combined 
    BAO+CMB+SNe data set. The dashed line indicates the Maximum A Posteriori (MAP) 
    value $\log_{10}c_s^2=-0.9780$ for the $w_0w_a$CDM model. 
    Right: Reconstructed evolution of the dark energy sound speed squared 
    $c_s^2(z)$ 
    within the EFT framework for $w$CDM (blue) and $w_0w_a$CDM (green). Solid lines 
    denote posterior mean values, while dashed and dot–dashed curves show the 68\% 
    and 95\% confidence intervals. The reconstruction is shown up to $z=5$, beyond 
    which $\Omega_{\rm de}$ becomes negligible.
    }}}
    \label{fig:compare_cs2_all}
\end{figure*}

\section{The EFT approach}
In addition to the fluid 
description adopted in the PPF framework, dark energy may also be interpreted 
as a scalar degree of freedom or as an effective manifestation of modified 
gravity. 
A unified and model-independent description of such scenarios is provided by 
the EFT of dark energy, which captures both background 
evolution and linear perturbations within a single formalism 
\cite{Gubitosi:2012hu,Bloomfield:2012ff,Creminelli:2008wc,Gleyzes:2013ooa, Li:2018ixg}. 
A key advantage of the EFT approach is that perturbations can be consistently 
analyzed independently of the assumed background expansion history.

Given the broad landscape of dark energy and modified-gravity theories, we 
adopt a parametric EFT description based on the $\alpha$-basis 
\cite{Bellini:2014fua,Hu:2014oga}, which efficiently captures the phenomenology 
of the Horndeski class of theories \cite{Horndeski:1974wa}. 
In this framework, departures from General Relativity are encoded in a small 
set of time-dependent functions $\alpha_{\rm M}$, $\alpha_{\rm B}$, 
$\alpha_{\rm K}$, and $\alpha_{\rm T}$. 
Current constraints on the speed of gravitational waves motivate setting 
$\alpha_{\rm T}=0$ throughout our analysis \cite{LIGOScientific:2017zic}. 

Consistency of the EFT description requires the absence of ghost and gradient 
instabilities, which imposes well-known theoretical conditions on the $\alpha$ 
functions 
\cite{Sbisa:2014pzo,Cline:2003gs,Carroll:2003st,Gumrukcuoglu:2016jbh,DeFelice:2016ucp}. 
Within this framework, the effective sound speed of scalar perturbations is 
determined by combinations of the $\alpha$ functions and the background 
quantities \cite{Bellini:2014fua}. 
In the present analysis, we enforce the physically viable range $0<c_s^2\leq 1$. See
Appendix.~B for details of the EFT approach.

Since the $\alpha$ functions primarily control the behavior of perturbations, 
a complete cosmological model additionally requires specifying the background 
expansion history. 
Accordingly, we consider both constant-$w$ and dynamical $w_0w_a$ backgrounds, 
denoted as $w$CDM+EFT and $w_0w_a$CDM+EFT, respectively. 
Following common practice, we parametrize the time dependence of the EFT 
functions as $\alpha_i(t)=c_i\,\Omega_{\rm de}(t)$, where $\Omega_{\rm de}(t)$ 
is the dark energy density parameter and $c_i$ are constant coefficients and $i \in \{M, B, K\}$
\cite{Pujolas:2011he, Barreira:2014jha, Bellini:2015oua}. 
This choice ensures that deviations from General Relativity become negligible 
at early times, while allowing for nontrivial dark energy dynamics at late 
epochs.\\



\section{Method and dataset}
We use the Monte Carlo Markov chain (MCMC)
analysis to constrain the perturbative properties of 
dynamical dark energy using the latest cosmological observations, with the MCMC setup detailed in Appendix.~C. 
Our data combination includes the full Planck 2018 CMB temperature and 
polarization spectra and CMB lensing reconstruction 
\cite{Planck:2018vyg}, the Union3 compilation of 2087 Type~Ia 
supernovae in the redshift range $0.05<z<2.26$, and baryon acoustic oscillation 
measurements from DESI Data Release~2 \cite{DESI:2025zgx}.

The clustering behavior of dark energy is governed by its effective sound speed 
$c_s^2$ through the comoving Jeans scale $k_{\rm J}^2\sim a^2H^2/c_s^2$. 
For $c_s^2\simeq1$, dark energy remains smooth on sub-horizon scales, while for 
$c_s^2\ll1$ it can cluster and affect the growth of matter perturbations and 
gravitational potentials. 
Since phenomenologically relevant values may be very small, following the 
Planck Collaboration \cite{Planck:2015bue} we parametrize the sound 
speed as $\log_{10}c_s^2$ within the PPF framework.\\

\begin{figure*}[htbp]
\centering
\includegraphics[width=\textwidth]{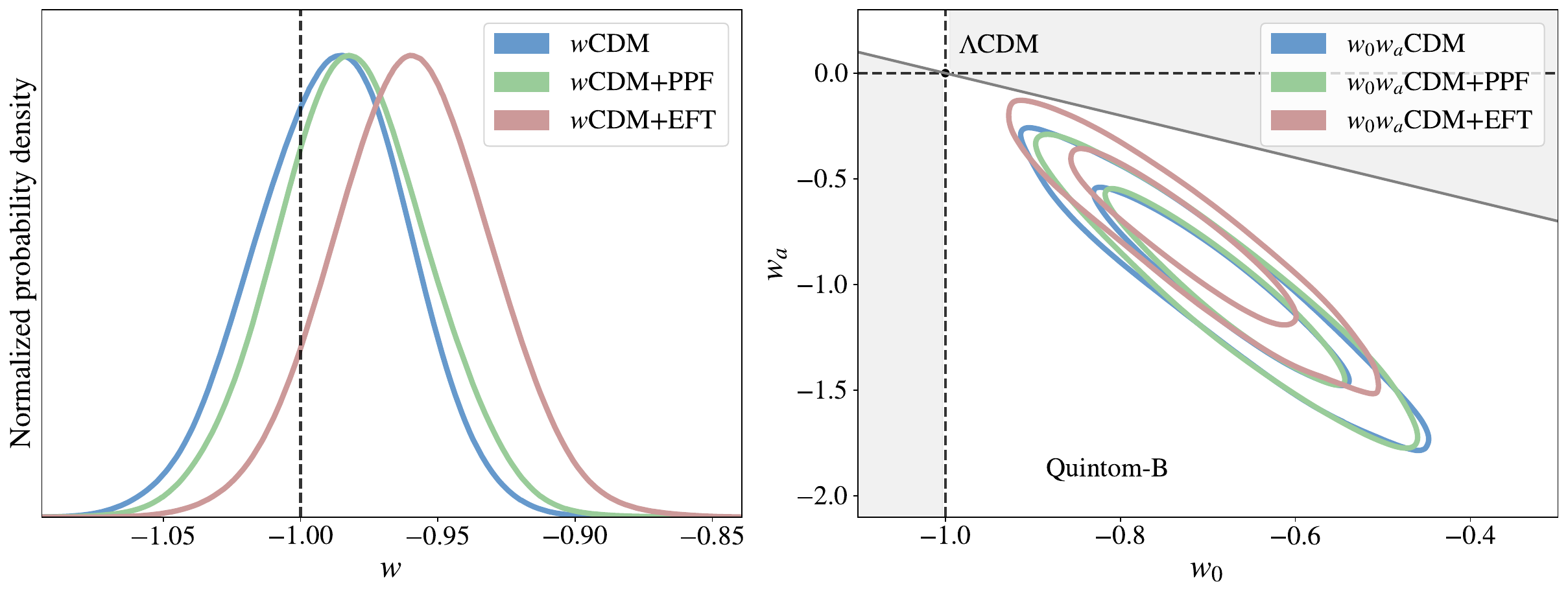}
\caption{{\it{Left: Normalized marginalized one–dimensional posterior 
distributions for the equation-of-state parameter $w$ in a $w$CDM background under different perturbative descriptions, obtained from the combined BAO+CMB+SNe data set. 
The dashed line denotes the $\Lambda$CDM limit $w=-1$. 
Right: Constraints in the $(w_0,w_a)$ plane for the $w_0w_a$CDM model under 
different perturbative descriptions. Contours enclose 68\% and 95\% confidence regions. The gray line corresponds to $w_0+w_a=-1$, while the $\Lambda$CDM point ($w_0=-1$, $w_a=0$) lies at the intersection of the axes. 
The deviation from $\Lambda$CDM reaches $3.42\sigma$ for $w_0w_a$CDM, increasing to $3.63\sigma$ when perturbations are treated within the PPF framework and decreasing to $3.19\sigma$ in the EFT case. 
}}}
    \label{fig:compare_w_all}
\end{figure*}


\section{Numerical results}
We now present the constraints on dark energy 
perturbations obtained from the combined BAO, CMB, and Type~Ia supernova data sets. 
Our analysis focuses on the effective sound speed of dark energy and its role in determining the clustering properties of dynamical models. 
The results are obtained within both the PPF and EFT perturbative frameworks, while technical details of the statistical analysis and additional consistency checks are deferred to the Appendix.~D.

Including dark energy perturbations within the PPF framework has only a minor 
impact on the inferred background cosmological parameters, whose central values 
shift slightly without any substantial change in their uncertainties. 
This behavior is expected, since the BAO and Type~Ia supernova data primarily 
constrain the background expansion history. 
Sensitivity to dark energy perturbations arises almost entirely from CMB 
observations, in particular from the large-scale temperature and polarization 
anisotropies. 
Owing to the sizable cosmic variance on these scales, current data do not 
tightly constrain the sound speed in general, nor do they exclude the smooth 
dark energy limit $c_s^2=1$ with high statistical significance.

The left panel of Fig.~\ref{fig:compare_cs2_all} shows the marginalized 
one-dimensional posterior distributions of $\log_{10}c_s^2$ obtained within the PPF approach, and we have considered the impact of hard prior bounds on the posterior distribution.
For the constant-$w$ case ($w$CDM+PPF), the sound speed remains essentially 
unconstrained. 
In contrast, when a dynamical equation-of-state parameter is allowed, the $w_0w_a$CDM+PPF model exhibits a clear preference for small sound-speed values, yielding
\begin{equation}
 \log_{10}c_s^2 = -3.00^{+2.9}_{-0.99} ~.
\end{equation}
To test the projection effects that may occur when nuisance
parameters are partially degenerate with
cosmological parameters, we maximize the log-posterior to get a    Maximum A Posteriori (MAP) value $\log_{10}c_s^2\simeq -0.98$, corresponding to $c_s^2\simeq0.1$. 
This result indicates that current observations are sensitive to the clustering properties of dark energy provided that its equation-of-state parameter evolves in time.

The physical origin of this behavior can be understood from the Poisson 
equation, where the contribution of dark energy perturbations scales as 
$(1+w_{\rm de})$. 
In constant-$w$ models close to $w=-1$, this factor strongly suppresses the 
impact of dark energy clustering, rendering the sound speed effectively 
unconstrained even for very small values. 
In the $w_0w_a$CDM scenario, however, the time dependence of the equation-of-state parameter partially lifts the degeneracy between $(1+w_{\rm de})$ and $c_s^2$, allowing meaningful constraints to be placed. 
We stress that this improvement is driven by the dynamical nature of the 
equation-of-state itself, rather than by the specific perturbative prescription, explaining why the $w_0w_a$CDM sound speed can be constrained whereas the $w$CDM one cannot.

To test the robustness of these findings, we perform a complementary analysis within the effective field theory (EFT) framework. 
The reconstructed evolution of the effective sound speed, shown in the right panel of Fig.~\ref{fig:compare_cs2_all}, displays similar behavior for both $w$CDM+EFT and $w_0w_a$CDM+EFT models, with mean values $c_s^2\simeq0.3$--$0.4$ and broad uncertainties. 
Since dark energy is negligible at early times, the reconstruction is shown only up to $z=5$. 
The asymptotic approach of $c_s^2$ at higher redshifts reflects the adopted 
parametrization $\alpha_i(z)\propto\Omega_{\rm de}(z)$, which recovers the standard quintessence-like behavior during matter domination \cite{Ratra:1987rm, Wetterich:1987fm}. It is important to note, that the physical meaning of the sound speed in the EFT framework differs from that in the fluid description. In the EFT, the sound speed is defined as the propagation speed of the scalar mode. The origin of this scalar degree of freedom can arise from various physical mechanisms, including additional scalar fields, non-minimally coupled matter, modified gravity, and others. Therefore, a distinction from the sound speed defined in the fluid dark energy model is necessary. In this context, the EFT can be understood as a modification to the physics at the perturbative level.


The implications of including dark energy perturbations for the background 
equation-of-state parameter are illustrated in Fig.~\ref{fig:compare_w_all}. 
The left panel shows the marginalized mean values and $68\%$ confidence 
intervals for constant-$w$ models under different perturbative treatments. 
The $w$CDM and $w$CDM+PPF results are nearly identical, indicating that the 
inclusion of fluid dark energy perturbations does not significantly affect the inferred equation-of-state in this case. 
In contrast, the $w$CDM+EFT scenario yields a slightly higher value of $w$, 
reflecting the additional freedom introduced by modified-gravity effects at the 
perturbative level.

The right panel of Fig.~\ref{fig:compare_w_all} displays the constraints in the 
$(w_0,w_a)$ plane for the $w_0w_a$CDM model. 
The dashed line corresponds to $w_0+w_a=-1$, separating non-phantom from 
phantom-crossing evolutions. 
For both the PPF and EFT approaches, the allowed parameter regions lie 
predominantly below this line, signaling a quintom-like (quintom-B) behavior of the dark energy equation-of-state, with phantom-like evolution at earlier times and quintessence-like behavior at late epochs \cite{Feng:2004ad,Hu:2004kh}. 

Quantitatively, we find a deviation from $\Lambda$CDM at the $3.42\sigma$ level 
for the $w_0w_a$CDM model. 
This significance increases mildly to $3.63\sigma$ when perturbations are 
treated within the PPF framework, and decreases to $3.19\sigma$ in the EFT 
case. 
The difference with respect to the significance reported for the same data 
combination in Ref.~\cite{DESI:2025zgx} arises from the use of different CMB 
likelihoods and lensing data sets, and does not alter the qualitative preference 
for dynamical dark energy.


To further assess the relative performance of the different models, we evaluate the standard information criteria in Table~\ref{table:aic_bic_result}, namely the Akaike Information Criterion (AIC) and 
the Bayesian Information Criterion (BIC) \cite{Liddle:2007fy,Anagnostopoulos:2019miu}. 
We find that models with a dynamical equation-of-state, such as $w_0w_a$CDM, are 
mildly favored by AIC, reflecting their improved fit to the data. 
This preference is slightly reduced when dark energy clustering is included, due 
to the increase in parameter space.

In contrast, the BIC continues to favor $\Lambda$CDM, as expected given its 
stronger penalty for model complexity. 
These results indicate that, while current data do not yet provide decisive 
evidence against the cosmological constant from a model-selection perspective, they do exhibit a statistically significant preference for dynamical dark energy and allow, for the first time, meaningful constraints on its perturbative properties.\\

\begin{table}[ht]
\centering
\caption{The information criteria AIC and BIC values as well as the corresponding differences $\Delta$AIC= $\mathrm{AIC_{model}}$ -$\mathrm{AIC_{\Lambda CDM}}$ and $\Delta$BIC= $\mathrm{BIC_{model}}$ -$\mathrm{BIC_{\Lambda CDM}}$, for different models. }
\begin{tabular}{lcccc}
\hline
Model & ~ AIC & ~ BIC & ~ $\Delta$AIC & ~ $\Delta$BIC \\
\hline

$\Lambda$CDM & ~ $2879.90$ & ~ $3041.49$ & ~ $0$ & ~ $0$ \\

$w$CDM & ~ $2882.18$ & ~ $3049.54$ & ~ $2.28$ & ~ $8.05$\\

$w$CDM+PPF & ~ $2882.00$ & ~ $3055.13$ & ~ $2.10$ & ~ $13.64$\\

$w$CDM+EFT & ~ $2878.84$ & ~ $3063.51$  & ~ $-1.06$ & ~ $22.02$\\

$w_0w_a$CDM & ~ $2870.54$ & ~ $3043.67$ & ~ $-9.36$ & ~ $2.18$\\

$w_0w_a$CDM+PPF & ~ $2872.16$ & ~ $3051.06$ & ~ $-7.71$ & ~ $9.57$\\

$w_0w_a$CDM+EFT & ~ $2870.82$ & ~ $3061.27$ & ~ $-9.08$ & ~ $19.78$\\
\hline

\end{tabular}
\label{table:aic_bic_result}
\end{table}


\section{Conclusions}
In this paper we investigated the perturbative 
properties of dark energy by probing its effective sound speed and clustering behavior using current cosmological observations. 
Motivated by recent DESI DR2 results favoring a dynamical equation-of-state, we performed a joint analysis within two complementary and theoretically consistent frameworks: the PPF approach, which provides a stable fluid description across the $w=-1$ boundary, and the EFT of dark energy, which offers a model-independent parametrization of perturbations.

Within the PPF framework, we focused on $\log_{10}c_s^2$ in order to access the phenomenologically relevant regime of very small sound speeds. 
For constant-$w$ models, the sound speed remains essentially unconstrained, as expected for dark energy close to a cosmological constant. 
In contrast, when a dynamical equation-of-state is allowed, the degeneracy 
between $(1+w)$ and $c_s^2$ is partially lifted, yielding the constraint
$\log_{10}c_s^2=-3.00^{+2.9}_{-0.99}$, with a MAP value 
$\log_{10}c_s^2\simeq-0.98$. 
This constitutes the first meaningful constraint on the sound speed of dynamical dark energy from current data. 
The EFT analysis provides a consistent and complementary picture, with 
reconstructed sound-speed values in the range $c_s^2\sim0.3$ or $0.4$.

We further quantified the deviation from $\Lambda$CDM in the $w_0w_a$CDM 
background, finding a $3.4\sigma$ preference for dynamical dark energy, mildly enhanced when perturbations are treated within the PPF framework and slightly reduced in the EFT case. 
While information criteria continue to favor $\Lambda$CDM due to its minimal parameter content, they nonetheless indicate that dynamical dark energy models provide an improved fit to the data.

While previous analyses, including Planck 2015, found the dark energy sound speed to be essentially unconstrained, our results successfully demonstrate that the combination of DESI DR2, CMB, and supernova data for the first time has reached the sensitivity required to probe the clustering properties of dark energy. The perturbative behavior of dark energy opens new avenues to test the physical origin of cosmic acceleration. Dark energy sound speed, like its equation-of-state, can become one of the cornerstones of modern cosmology, providing a novel probe into the cosmic acceleration by linking background expansion and perturbation evolution. 
Recent hydrodynamical simulations \cite{Blot:2026var} reveal that clustering dark energy develops significant nonlinear perturbations on small scales, contributing $\sim 10\%$ to the total density within massive dark matter halos ($\delta_{\text{de}} \approx 50$ at the center). This alters the matter power spectrum at $z < 1$ and can be probed observationally via stacked gravitational lensing and dynamical measurements.
Future surveys, such as the full DESI data release, Euclid, LSST, CMB-S4, and SKA, will decisively test whether this emerging sensitivity reveals genuine clustering of dark energy or confirms its smooth behavior, thereby shedding light on the destiny of the universe. \\



\section*{Conflict of interest}
The authors declare that they have no conflict of interest.

\begin{acknowledgments}
We are grateful to Linda Blot, Hao Liu and Zhiyu Lu for valuable comments. This work is supported in part by the National Key R\&D Program of China 
(2021YFC2203100), by the NSFC (12433002, 12261131497, 92476203), by CAS young interdisciplinary innovation team (JCTD-2022-20), by 111 Project (B23042), by Anhui Postdoctoral Scientific Research Program Foundation (No. 2025C1184), by CSC Innovation Talent Funds, by USTC Fellowship for International Cooperation, and by USTC Research Funds of the Double First-Class Initiative. ENS acknowledges the contribution of the LISA Cosmology Working Group (CosWG). They acknowledge as well support from the COST Actions CA21136 -  Addressing observational tensions in cosmology with systematics and fundamental physics (CosmoVerse) - CA23130, Bridging high and low energies in search of quantum gravity (BridgeQG) and CA21106 - COSMIC WISPers in the Dark Universe: Theory, astrophysics and  experiments (CosmicWISPers). Kavli IPMU is supported by World Premier International Research Center 
Initiative (WPI), MEXT, Japan. XR is supported by ``Talent Scientific Fund of Lanzhou University'' and the activity ``APCTP-2026-F02''.
\end{acknowledgments}

\section*{Author contributions}
Yi-Fu Cai conceived the idea. He also initiated this study with all other authors. Yuhang Yang conducted numerical calculations and analyzed physical results. Yuhang Yang, Qingqing Wang, Xin Ren, Yi-Fu Cai, and Emmanuel N. Saridakis wrote the manuscript. Emmanuel N. Saridakis provided many valuable suggestions on this work. All authors discussed the results together.



\begin{widetext}
\appendix

\section{Parameterized Post-Friedmann (PPF) implementation}
\label{appendix:ppf}

Within the Parameterized Post-Friedmann (PPF) framework, dark energy is treated 
as an additional non-interacting fluid alongside cold dark matter and 
radiation, 
with equation of state $w=p_{\rm de}/\rho_{\rm de}$. At linear order, its 
perturbations are described by the density contrast 
$\delta_{\rm de}\equiv\delta\rho_{\rm de}/\rho_{\rm de}$ and the velocity 
divergence $\theta_{\rm de}\equiv\partial_i v_{\rm de}^i$, while anisotropic 
stress is assumed to vanish.

The PPF formalism introduces a single auxiliary variable $\Gamma$ that governs 
the evolution of dark-energy perturbations while preserving local 
energy--momentum conservation \cite{Fang:2008sn,Hu:2008zd}. The relevant 
definitions are
\begin{align}
\Gamma &\equiv -\frac{4\pi G a^2}{k^2\mu_K}\rho_{\rm de}
\Delta_{\rm de}^{\rm rest}, \\
S &\equiv -\frac{4\pi G}{H^2}
\Big[f_\zeta(t)(\rho_{\rm T}+p_{\rm T})-\rho_{\rm de}(1+w)\Big]
\frac{v_{\rm T}+k\alpha}{k_H},
\end{align}
where $\Delta_{\rm de}^{\rm rest}$ denotes the dark-energy density contrast in 
its rest frame and the subscript ``T'' refers to all components other than dark 
energy. Here $\alpha=a(\dot{h}+6\dot{\eta})/(2k^2)$, with $h$ and $\eta$ the 
metric 
perturbations in synchronous gauge \cite{Ma:1995ey}, $k_H\equiv k/(aH)$, and 
$\mu_K=1-3K/k^2$, with $K$ the background curvature. In this work we set 
$\mu_K=1$ and adopt $f_\zeta=0$, which has been shown to be sufficient for most 
cosmological applications \cite{Fang:2008sn}.

Dark-energy perturbations are assumed to become smooth relative to matter below 
a transition scale defined by $c_s k_H=1$. Their evolution is governed by
\begin{equation}
\left(1+c_\Gamma^2 k_H^2\right)
\left(\frac{\dot{\Gamma}}{H}+\Gamma+c_\Gamma^2 k_H^2\Gamma\right)=S,
\end{equation}
where we take $c_\Gamma=0.4\,c_s$ following Ref.~\cite{Fang:2008sn}. This 
prescription ensures a regular and stable evolution of perturbations across the 
$w=-1$ boundary. In the large-scale limit $k_H\rightarrow0$, the full system of 
perturbation equations can be consistently solved, yielding well-defined metric 
and gravitational potential evolution. The impact of dark-energy perturbations 
on large-scale observables, including the integrated Sachs--Wolfe effect, is 
illustrated in Fig.~\ref{fig:cmb_isw}.

\begin{figure}[htbp]
    \includegraphics[width=\textwidth]{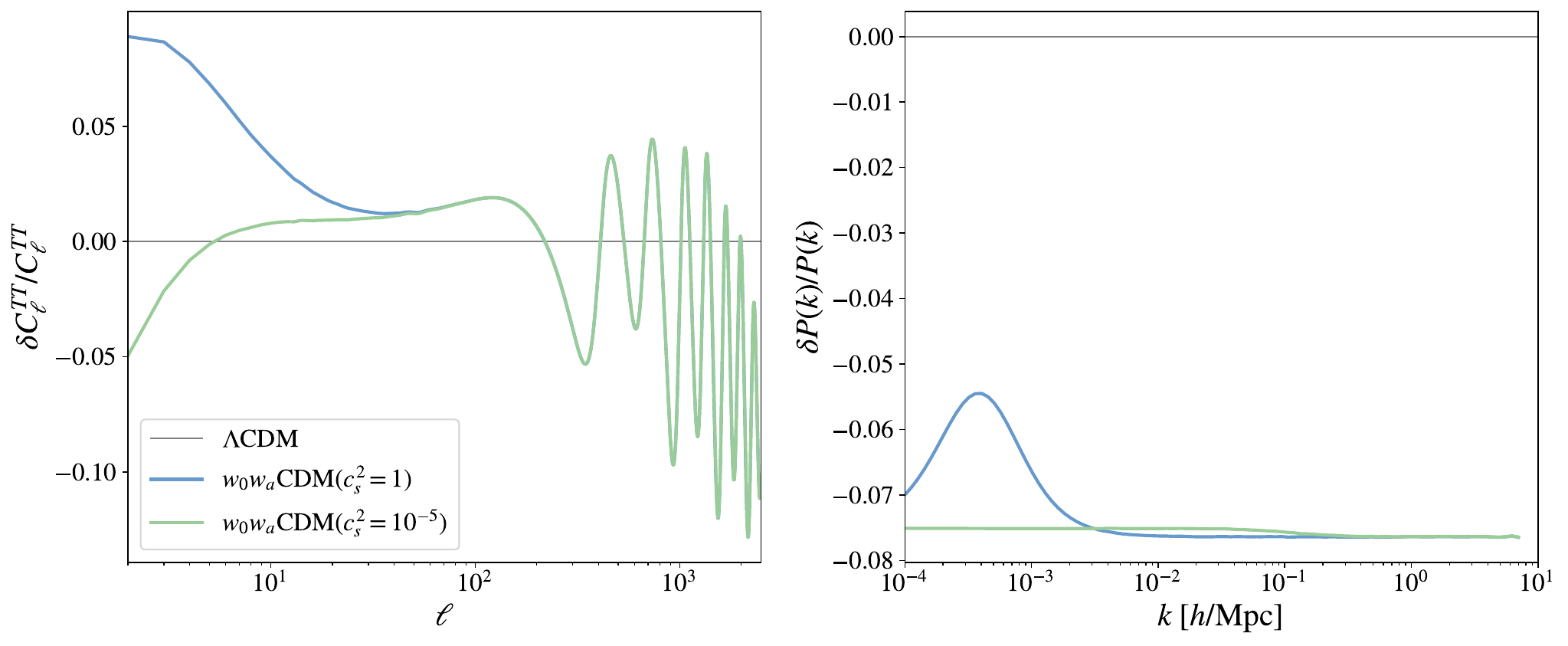}
    
    \caption{\it{Residuals of CMB temperature power spectrum (left) and matter power spectrum 
(right) for $\Lambda$CDM model and the 
$w_0w_a$CDM model with different values of the dark energy sound speed within the PPF framework. The black solid line 
represents the $\Lambda$CDM model, the blue and green solid lines represent the smooth
$w_0w_a$CDM model and clustering $w_0w_a$CDM model, 
respectively.}}
    \label{fig:cmb_isw}
\end{figure}

\section{Effective field theory (EFT) implementation}
\label{appendix:eft}

We adopt the effective field theory (EFT) of dark energy in unitary gauge, in 
which the most general action describing a single scalar degree of freedom 
coupled to gravity can be written as \cite{Gubitosi:2012hu,Bloomfield:2012ff,
Creminelli:2008wc,Park:2010cw}
\begin{equation}
\begin{aligned}
S = \int d^4x \sqrt{-g} \Big[ &
\frac{M_{\rm P}^2}{2}\big(1+\Omega(t)\big)R
-\Lambda(t)-c(t)g^{00}
+\frac{M_2^4(t)}{2}(\delta g^{00})^2
-\frac{\bar M_1^3(t)}{2}\delta g^{00}\delta K \\
&-\frac{\bar M_2^2(t)}{2}(\delta K)^2
-\frac{\bar M_3^2(t)}{2}\delta K^\mu_{\ \nu}\delta K^\nu_{\ \mu}
+\frac{\hat M^2(t)}{2}\delta g^{00}\delta R^{(3)}
+m_2(t)\partial_i g^{00}\partial^i g^{00}
+\cdots \Big]  \\
&+ S_{\rm m}[g_{\mu\nu},\Psi_{\rm m}],
\end{aligned}
\label{eq:eft_action}
\end{equation}
where $\delta g^{00}\equiv g^{00}+1$, $\delta K^\mu_{\ \nu}$ is the 
perturbation 
of 
the extrinsic curvature with trace $\delta K$, and $\delta R^{(3)}$ denotes the 
perturbation of the spatial Ricci scalar. The matter action 
$S_{\rm m}[g_{\mu\nu},\Psi_{\rm m}]$ includes all non–dark-energy species, 
which 
are assumed to satisfy the weak equivalence principle and are minimally coupled 
to the metric (Jordan frame).

The time-dependent functions $\Lambda(t)$ and $c(t)$ determine the background 
evolution, while $\Omega(t)$ parametrizes a nonminimal coupling to gravity. The 
remaining operators primarily affect the dynamics of perturbations. Their 
specific form is fixed once a covariant theory underlying the EFT description 
is 
specified.

For phenomenological analyses we employ the $\alpha$-basis, which captures the 
linear perturbation dynamics of the Horndeski class of theories 
\cite{Bellini:2014fua,Hu:2014oga}. Deviations from General Relativity are 
encoded 
in four time-dependent functions,
\begin{equation}
\begin{aligned}
\alpha_{\rm M} &\equiv \frac{d\ln M_*^2}{d\ln a}, \qquad
\alpha_{\rm B} \equiv -\frac{M_{\rm P}^2\dot\Omega+\bar M_1^3}{H M_*^2}, \\
\alpha_{\rm K} &\equiv \frac{2c+4M_2^4}{H^2 M_*^2}, \qquad
\alpha_{\rm T} \equiv c_\mathrm{T}^2-1 = \frac{M_3^2}{M_*^2},
\end{aligned}
\end{equation}
where $M_*^2(t)=M_{\rm P}^2[1+\Omega(t)]-\bar M_3^2$ is the effective Planck 
mass 
and $c_\mathrm{T}$ denotes the propagation speed of tensor modes. Following the joint 
detection of GW170817 and GRB170817A \cite{LIGOScientific:2017zic}, we set 
$\alpha_{\rm T}=0$ throughout. The second-order action for perturbations of the tensors mode $h_{ij}$, and the scalar mode $\zeta$ is \cite{DeFelice:2011bh,Bellini:2014fua}
\begin{equation}
    S^{(2)}=\int\mathrm{d}t\mathrm{d^3}xa^3\Big [Q_S\Big( \dot{\zeta}^2 -\frac{c_s^2}{a^2} (\partial_i\zeta)^2 \Big) + Q_T \Big (\dot{h}_{ij}^2-\frac{c_\mathrm{T}^2}{a^2} (\partial_k h_{ij})^2 \Big ) \Big ],
\end{equation}
the stability of the scalar modes require
\begin{align}
        Q_s =&\frac{2M_*^2D}{(2-\alpha_\mathrm{B})^2}>0, \quad \quad D\equiv \alpha_\mathrm{K}+\frac{3}{2}\alpha_{\mathrm{B}}^2, \\
        c_s^2=& \frac{1}{H^2D}\Bigg \{ (\alpha_\mathrm{B}-2) \Big[\Dot{H}-\frac{1}{2}H^2\alpha_\mathrm{B}(1+\alpha_\mathrm{T})-H^2(\alpha_\mathrm{M}-\alpha_\mathrm{T})\Big] -H\Dot{\alpha}_{\mathrm{B}}+\rho_\mathrm{m}+p_\mathrm{m} \Bigg \}>0, \label{eq:cs2_eft}
\end{align}
while the stability of the tensor modes require
\begin{equation}
        Q_\mathrm{T}=\frac{M_*^2}{8}>0, \quad        c_\mathrm{T}^2=1+\alpha_{\mathrm{T}}>0.
\end{equation}
Taking all the conditions together also implies that the no-ghost condition for the scalar perturbations simply reduces to $D= \alpha_\mathrm{K}+\frac{3}{2}\alpha_{\mathrm{B}}^2> 0$. Meanwhile, the sound speed $c_s^2$ in the EFT framework is not fully equivalent with the one defined in the fluid approach, since the EFT framework can be also understood as a realization of modified gravity effects.

In the conformal Newtonian gauge, scalar perturbations of a flat 
Friedmann–Lemaître–Robertson–Walker background are described by
\begin{equation}
ds^2 = a^2(\tau)\big[-(1+2\Psi)d\tau^2 + (1-2\Phi)\delta_{ij}dx^i dx^j\big],
\end{equation}
where $\Psi$ and $\Phi$ are the gravitational potentials. Deviations from 
General 
Relativity can be conveniently characterized by the functions 
$\mu(a,k)=G_{\rm eff}/G$ and $\eta(a,k)=\Phi/\Psi$, or equivalently by 
$\Sigma=\mu(1+\eta)/2$, which govern the response of massive and massless 
particles, respectively. The modified Poisson equations take the form
\begin{align}
k^2\Psi &= -4\pi G a^2 \mu(a,k)\sum_i \rho_i \Delta_i^{\rm rest}, \\
k^2(\Phi+\Psi) &= -8\pi G a^2 \Sigma(a,k)\sum_i \rho_i \Delta_i^{\rm rest},
\end{align}
where $\Delta_i^{\rm rest}$ denotes the rest-frame density contrast of species 
$i$.

Within the $\alpha$-basis and assuming $\alpha_{\rm T}=0$, these 
functions are given by \cite{Bellini:2014fua,Ishak:2019aay}
\begin{align}
\mu(z) &= \frac{M_{\rm P}^2}{M_*^2}
\left[1+\frac{2(\alpha_{\rm M}+\tfrac{1}{2}\alpha_{\rm B})^2}
{c_s^2(\alpha_{\rm K}+\tfrac{3}{2}\alpha_{\rm B}^2)}\right],
\label{eq:MG_parameter_mu} \\
\Sigma(z) &= \frac{M_{\rm P}^2}{M_*^2}
\left[1+\frac{(\alpha_{\rm M}+\tfrac{1}{2}\alpha_{\rm B})
(\alpha_{\rm M}+\alpha_{\rm B})}
{c_s^2(\alpha_{\rm K}+\tfrac{3}{2}\alpha_{\rm B}^2)}\right],
\label{eq:MG_parameter_sigma}
\end{align}
which are the expressions used in our numerical implementation. And according to the stability conditions for scalar and tensor perturbations, the denominator is positively defined.


\section{Numerical implementation and MCMC analysis}
\label{appendix:mcmc}

We compute the background and linear perturbation evolution of the cosmological 
models using modified versions of 
\texttt{CLASS}~\cite{Blas2011TheCL} and 
\texttt{hi\_class}~\cite{Bellini:2019syt,Zumalacarregui:2016pph}. 
Parameter estimation is performed through a Monte Carlo Markov chain (MCMC) 
analysis with the public sampler \texttt{MontePython}~\cite{Audren:2012wb,
Brinckmann:2018cvx}. 
Convergence is assessed using the Gelman--Rubin diagnostic, requiring 
$R-1<0.02$ for all chains.

All analyses share a common set of baseline cosmological parameters: the 
physical 
densities of cold dark matter and baryons, 
$\omega_{\rm c}\equiv\Omega_{\rm c}h^2$ and 
$\omega_{\rm b}\equiv\Omega_{\rm b}h^2$; the approximate angular acoustic scale 
$\theta_{\rm MC}$; the amplitude $A_s$ and spectral index $n_s$ of the 
primordial 
curvature power spectrum; and the reionization optical depth $\tau$. 
Flat, uninformative priors are adopted for these parameters, as summarized in 
Table~\ref{table:mcmc_para}, which also lists the additional parameters 
specific 
to each dark-energy model. The priors for the baseline and dark-energy 
parameters follow Ref.~\cite{DESI:2024mwx}. Neutrinos are modeled as two 
massless 
and one massive species with $m_\nu=0.06$~eV, corresponding to 
$N_{\rm eff}=3.044$~\cite{Planck:2018vyg}.

\begin{table}[ht]
\centering
\caption{Parameters and priors used in the analysis. All     priors are flat in 
the ranges given.}
\begin{tabular}{cccc}
\hline
\textbf{Model} &\textbf{Parameter} & \textbf{Default} & \textbf{Prior}\\
\hline

Base &$\omega_\mathrm{b}$ & --- & $\mathcal{U}[0.005, 0.1]$ \\
&$\omega_\mathrm{cdm}$ & --- & $\mathcal{U}[0.001, 0.99]$  \\
&$100\theta_\mathrm{MC}$ & --- & $\mathcal{U}[0.5, 10]$  \\
&$\ln(10^{10}A_s)$ & --- & $\mathcal{U}[1.61, 3.91]$  \\
&$n_s$ & --- & $\mathcal{U}[0.8, 1.2]$  \\
&$\tau$ & --- & $\mathcal{U}[0.01, 0.8]$  \\
&$M$ & 0 & $\mathcal{U}[-5, 5]$  \\
\hline
DE&$w_0$ or $w$ & $-1$ & $\mathcal{U}[-3, 1]$ \\
&$w_a$ & 0 & $\mathcal{U}[-3, 2]$ \\
\hline
PPF&$\log_{10}c_s^2$ & 0 & $\mathcal{U}[-8, 0]$  \\
\hline
EFT&$c_\mathrm{K}$ & 0 & $\mathcal{U}[-10, 10]$  \\
&$c_\mathrm{B}$ & 0 & $\mathcal{U}[-10, 10]$  \\
&$c_\mathrm{M}$ & 0 & $\mathcal{U}[-10, 10]$  \\
\hline
\end{tabular}
\label{table:mcmc_para}
\end{table}

In addition to cosmological parameters, we sample the recommended nuisance 
parameters associated with each likelihood. Dark-energy perturbations are 
treated either within the PPF framework or within the EFT approach, as 
described 
in the previous sections. For the PPF analysis, and following the Planck 
Collaboration~\cite{Planck:2015bue}, we constrain $\log_{10}c_s^2$ rather than 
$c_s^2$ in order to properly explore the regime of very small sound speeds.

Our joint dataset consists of the following components:

\paragraph*{CMB.}
We use the full Planck 2018 cosmic microwave background 
data~\cite{Planck:2018vyg}, 
including temperature (TT), polarization (EE), and cross-correlation (TE) power 
spectra. The \texttt{Commander} and \texttt{SimAll} likelihoods are employed at 
low multipoles, while \texttt{Plik} is used for 
$\ell\geq30$~\cite{Planck:2019nip}. 
Planck CMB lensing measurements are also included~\cite{Planck:2018lbu}.

\paragraph*{Type Ia supernovae.}
We incorporate the Union3 compilation of 2087 Type Ia 
supernovae~\cite{Rubin:2023jdq}, 
covering the redshift range $0.05<z<2.26$ and standardized within the Unity~1.5 
Bayesian framework. The absolute magnitude $M$ is treated as a nuisance 
parameter, effectively encoding an offset relative to the calibrated baseline.

\paragraph*{BAO.}
We include baryon acoustic oscillation measurements from the DESI Data 
Release~2~\cite{DESI:2025zgx}, spanning a wide redshift range and multiple 
tracers: Bright Galaxy Sample ($0.1<z<0.4$), Luminous Red Galaxies 
($0.4<z<1.1$), Emission Line Galaxies ($1.1<z<1.6$), Quasars ($0.8<z<2.1$), 
and the Lyman-$\alpha$ forest ($1.77<z<4.16$).

\section{Supplementary constraints and MCMC results}
\label{appendix:result}

Figure~\ref{fig:eft_2d} shows the marginalized posterior distributions of the 
EFT 
parameters $c_{\rm M}$ and $c_{\rm B}$ for the $w$CDM+EFT and $w_0w_a$CDM+EFT 
models. The region $c_{\rm M}<0$ and $c_{\rm B}<0$ is excluded by gradient 
stability requirements. The data mildly prefer a nonzero braiding parameter 
$\alpha_{\rm B}$, while remaining consistent with vanishing Planck-mass running 
($\alpha_{\rm M}=0$), corresponding to minimally coupled gravity. The 
constraints 
are primarily driven by late-time effects on the metric potentials, in 
particular through the integrated Sachs--Wolfe contribution. The allowed region 
typically favors $c_{\rm B}>0$ and $c_{\rm M}<2$, consistent with expectations 
from CMB–galaxy cross-correlation analyses 
\cite{Stolzner:2017ged,Seraille:2024beb,Ishak:2024jhs,Lu:2025sjg}. We find that 
variations in the background expansion history have a negligible impact on the 
constraints of EFT parameters.

Using Eqs.~(\ref{eq:MG_parameter_mu}) and~(\ref{eq:MG_parameter_sigma}), we 
reconstruct the redshift evolution of the modified-gravity parameters $\mu$ and 
$\Sigma$, shown in Fig.~\ref{fig:eft_MG_parameter}. No statistically 
significant 
deviations from General Relativity are detected, in agreement with the absence 
of Planck-mass running and consistent with a minimally coupled dark-energy 
sector. 

In Fig.~\ref{fig:cmb_model_prediction}, we plot the comparison of CMB TT power spectrum between the model predictions and the \textit{Planck} observed. It can be seen that the main differences among various models mainly lie in the low-$\ell$ multipoles and the height of the first acoustic peak. For completeness, we present the full marginalized parameter constraints for 
all cosmological models and perturbation schemes considered in this work. The 
posterior distributions for the $w$CDM and $w_0w_a$CDM models, within both the 
PPF and EFT frameworks and using the combined BAO+CMB+SNe dataset, are shown in 
Figs.~\ref{fig:mcmc_all_wcdm}–\ref{fig:mcmc_all_w0wa}. A summary of the 
corresponding constraints is provided in Table~\ref{table:mcmc_result}.

From Table~\ref{table:mcmc_result}, we find that the kineticity parameter 
$c_{\rm K}$ remains weakly constrained in all cases, consistent with previous 
studies 
\cite{Bellini:2015xja,Reischke:2018ooh,Gleyzes:2015rua,Alonso:2016suf,
Yang:2018qmz,Pan:2019jqh}. As $c_{\rm K}$ has only a limited impact on 
background 
and large-scale observables \cite{Bellini:2015xja}, we keep it as a free 
parameter to ensure compliance with the physical stability condition 
$0<c_s^2\le1$.

\begin{figure} 
    \centering
    \centering
        \includegraphics[width=0.6\textwidth]{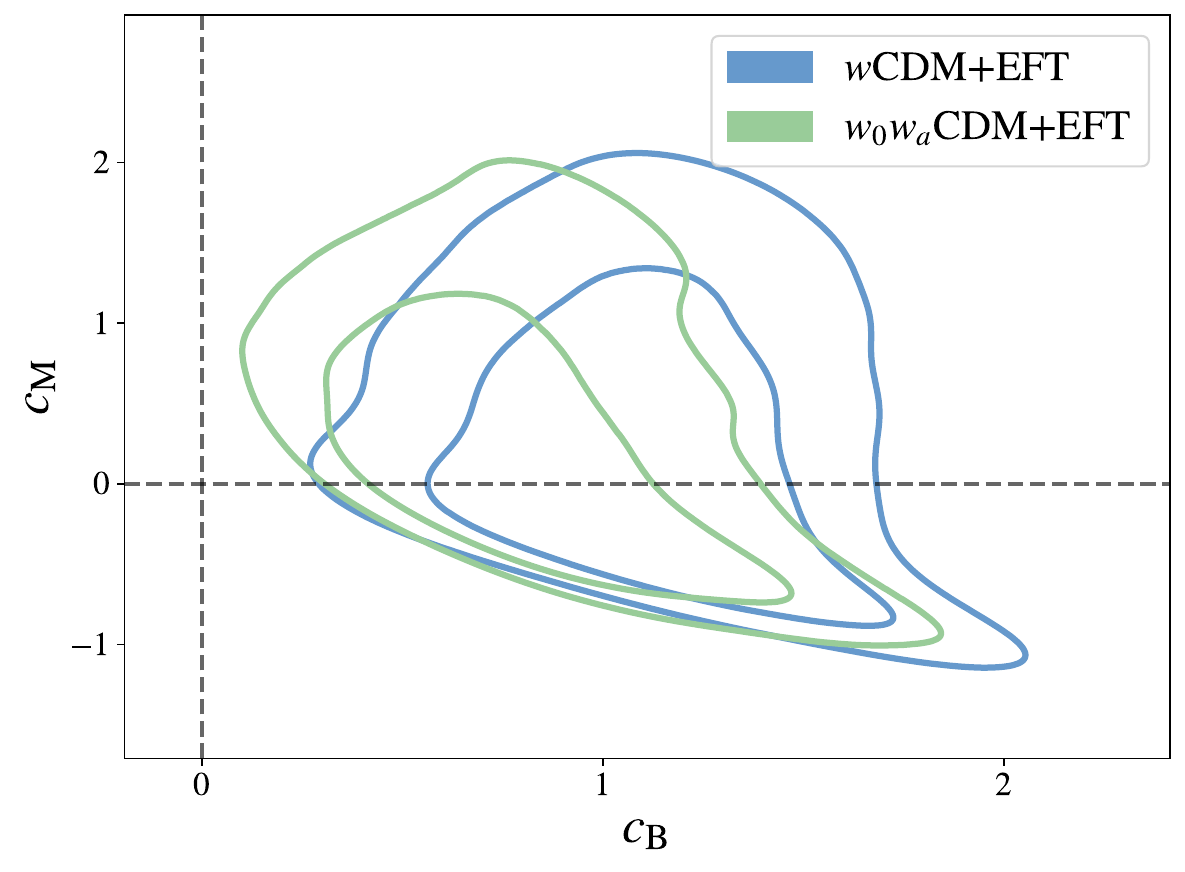}
    \caption{{\it{2D posterior distributions for the EFT coefficients 
$c_\mathrm{M}$ and $c_\mathrm{B}$, 
    for $w$CDM+EFT and $w_0w_a$CDM+EFT   using BAO+CMB+SNe datasets.}}}
    \label{fig:eft_2d}
\end{figure}

\begin{figure} 
    \centering
        \includegraphics[width=\textwidth]{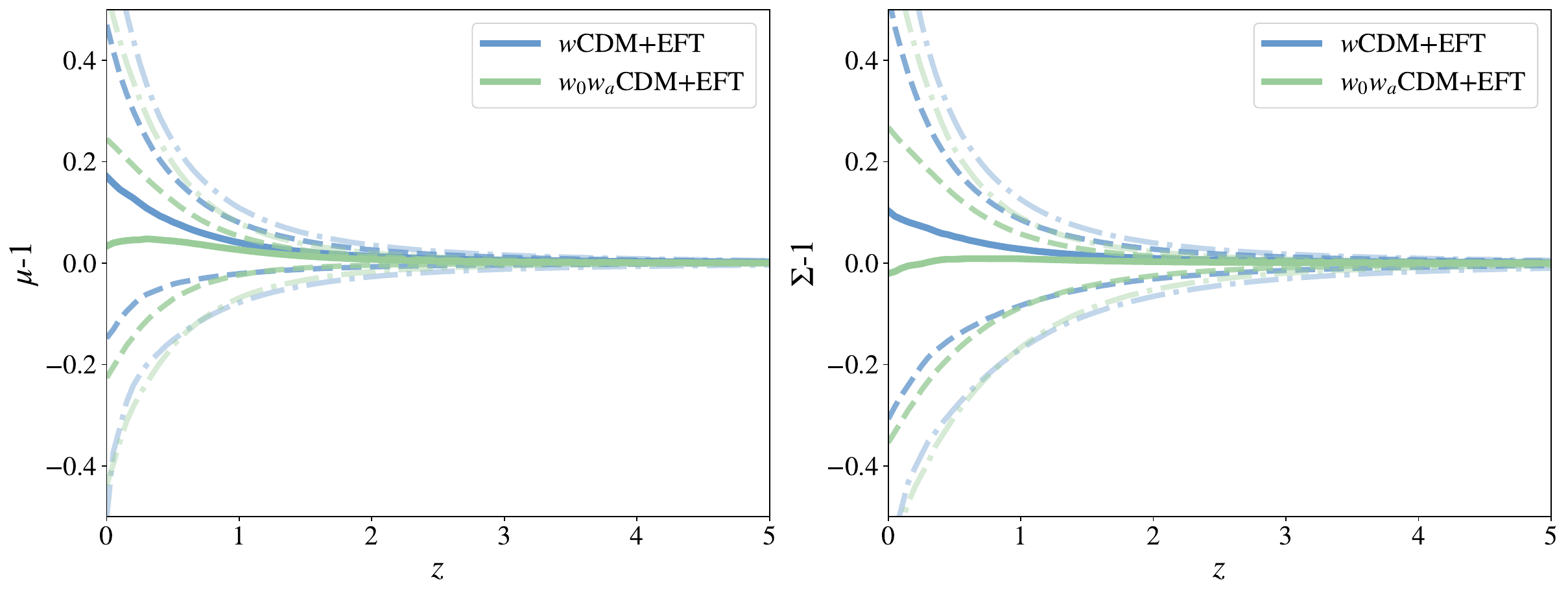}
    
    \caption{{\it{Left panel: Comparison of the reconstructed $\mu-1$  
    between $w$CDM+EFT (blue) and $w_0w_a$CDM+EFT (red), using BAO+CMB+SNe 
datasets. The solid line represents the mean value, and the dashed curve and 
dot-dashed curve
    represent 68\% and 95\% confidence intervals. Finally, the reconstruction is 
truncated at $z=5$ since $\Omega_\mathrm{de}$ becomes negligible at higher 
redshift. 
    Right panel:  Comparison of the reconstructed $\Sigma-1$  
    between $w$CDM+EFT (blue) and $w_0w_a$CDM+EFT (red), using BAO+CMB+SNe 
datasets.}}}
    \label{fig:eft_MG_parameter}
\end{figure}

\begin{figure}[htbp]
    \centering
    \begin{minipage}{0.49\textwidth}
        \includegraphics[width=\textwidth]{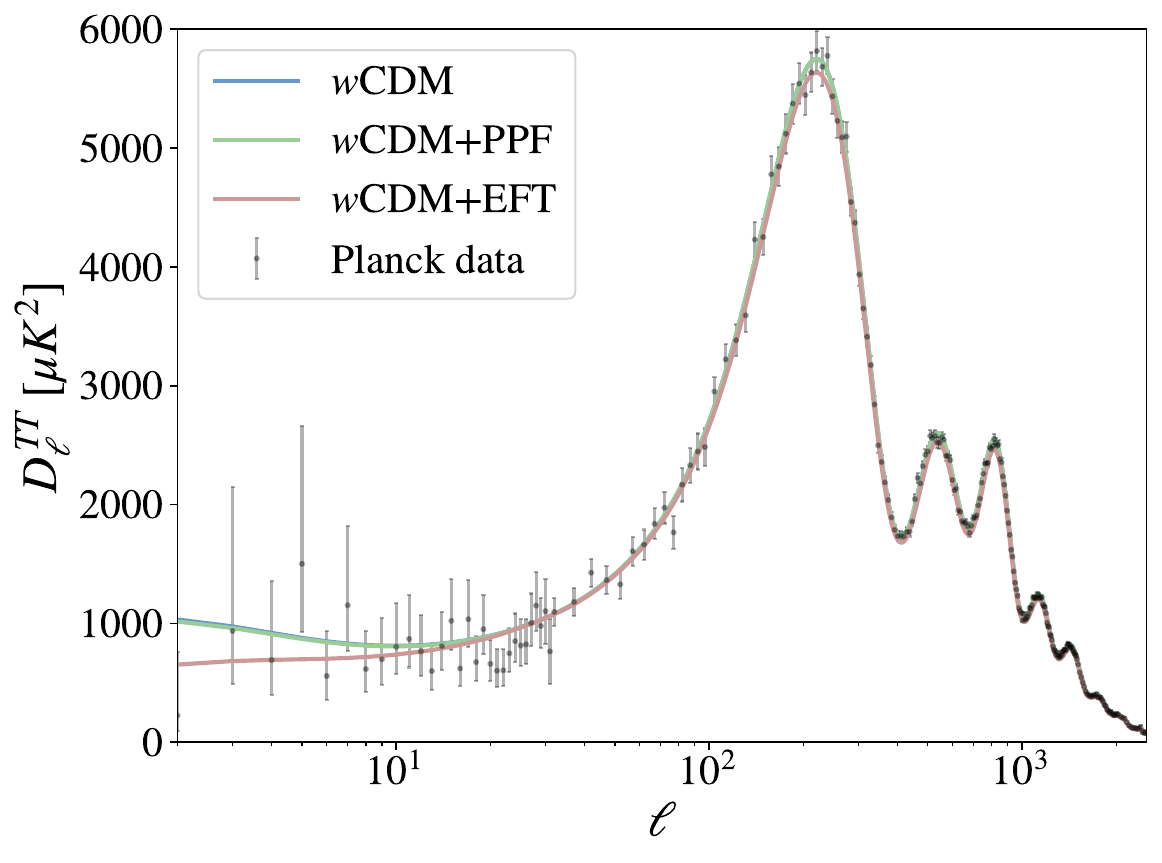}
    \end{minipage}
    \hfill
    \begin{minipage}{0.49\textwidth}
        \includegraphics[width=\textwidth]{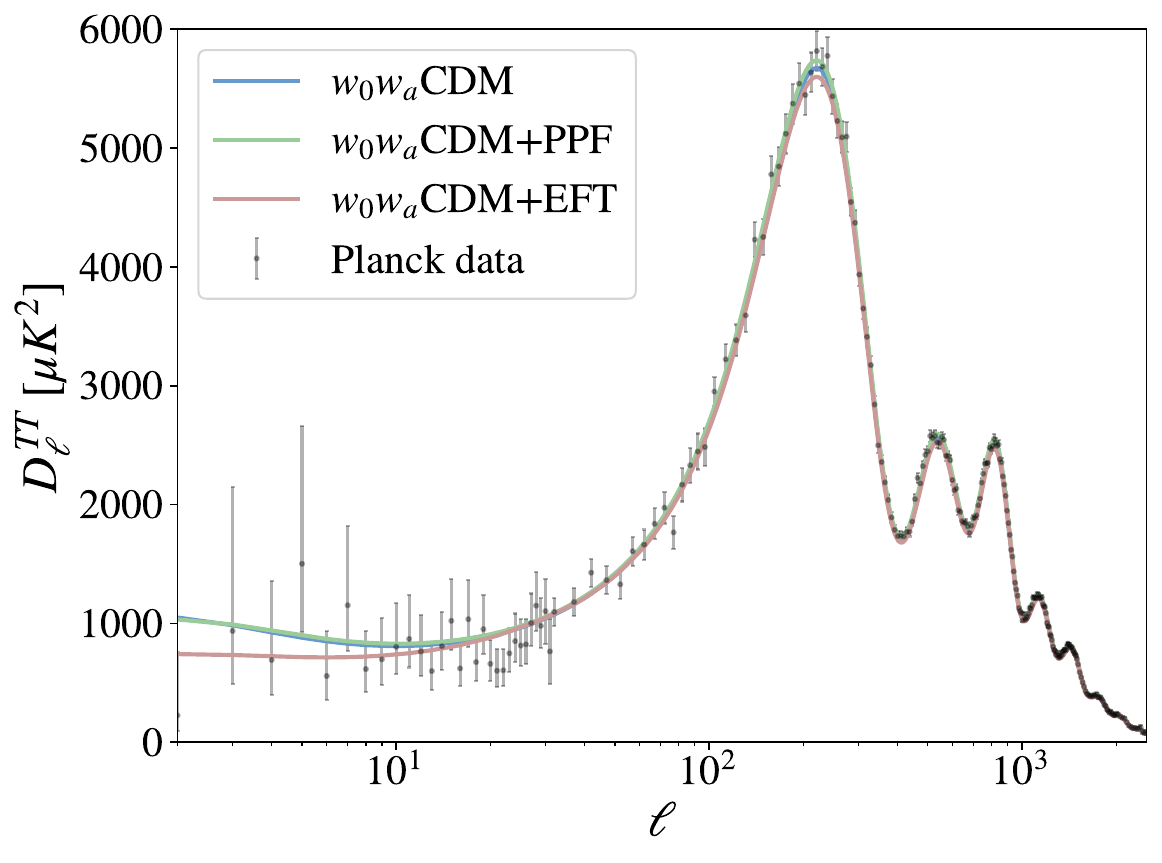}
    \end{minipage}
    
    \caption{\it{At multipoles $\ell \geq 30$ we show the temperature spectrum computed from the high-$\ell$ Plik likelihood, with foreground and other nuisance parameters fixed to a best fit assuming the base-$\Lambda$CDM cosmology. In the multipoles range $2 \le \ell \le 29$, we plot the power spectrum estimates from the low-$\ell$ likelihood Commander, computed over 86\% of the sky. And we also show the best-fit results for our constant-$w$ and $w_0w_a$ model with different perturbation descriptions.}}
    \label{fig:cmb_model_prediction}
\end{figure}
 
\begin{figure} 
    \centering
        
\includegraphics[width=\textwidth]{
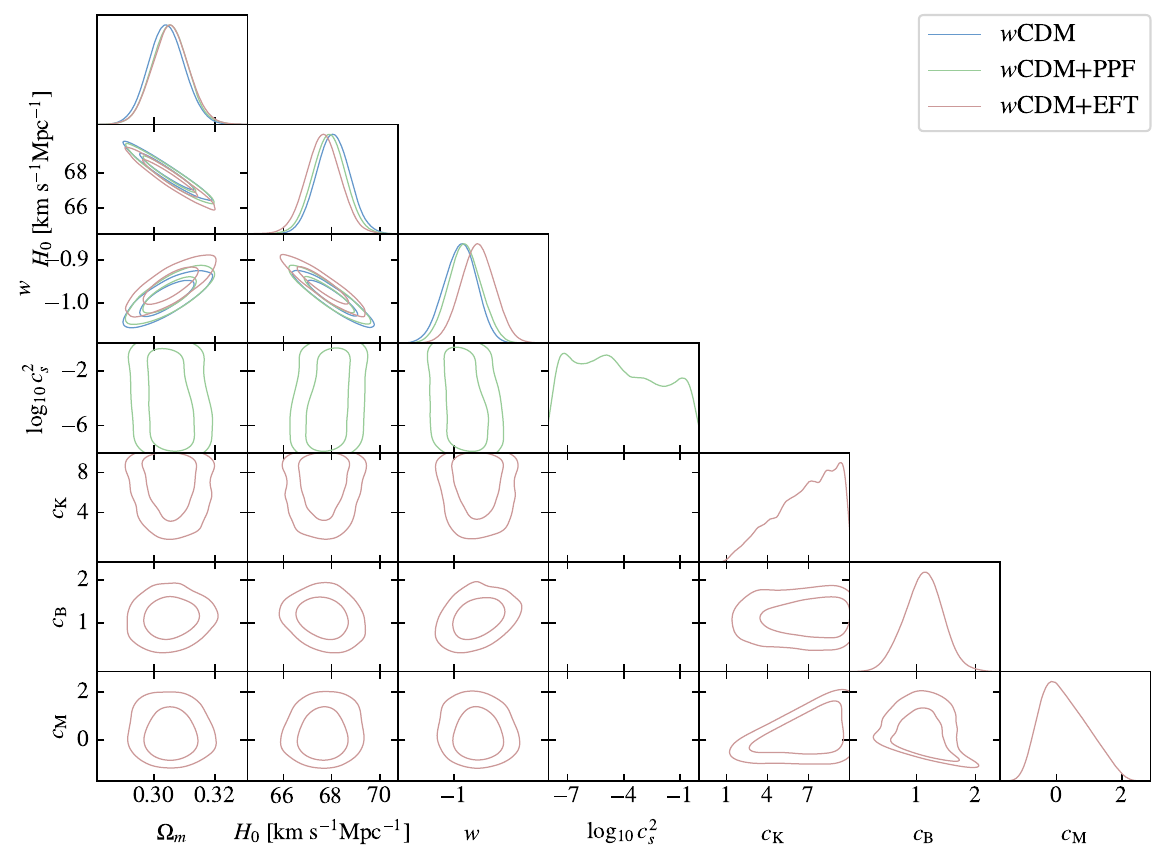}
    \caption{{\it{ Constraints on $w$CDM model under  Parameterized 
Post-Friedmann (PPF) and Effective Field Theory (EFT) perturbation descriptions, 
using BAO+CMB+SNe datasets. The contours represent the 68\% and 95\% credible 
intervals.  The results are summarized in Table~\ref{table:mcmc_result}.}}}
    \label{fig:mcmc_all_wcdm}
\end{figure}

\begin{figure} 
    \centering
        
\includegraphics[width=\textwidth]{
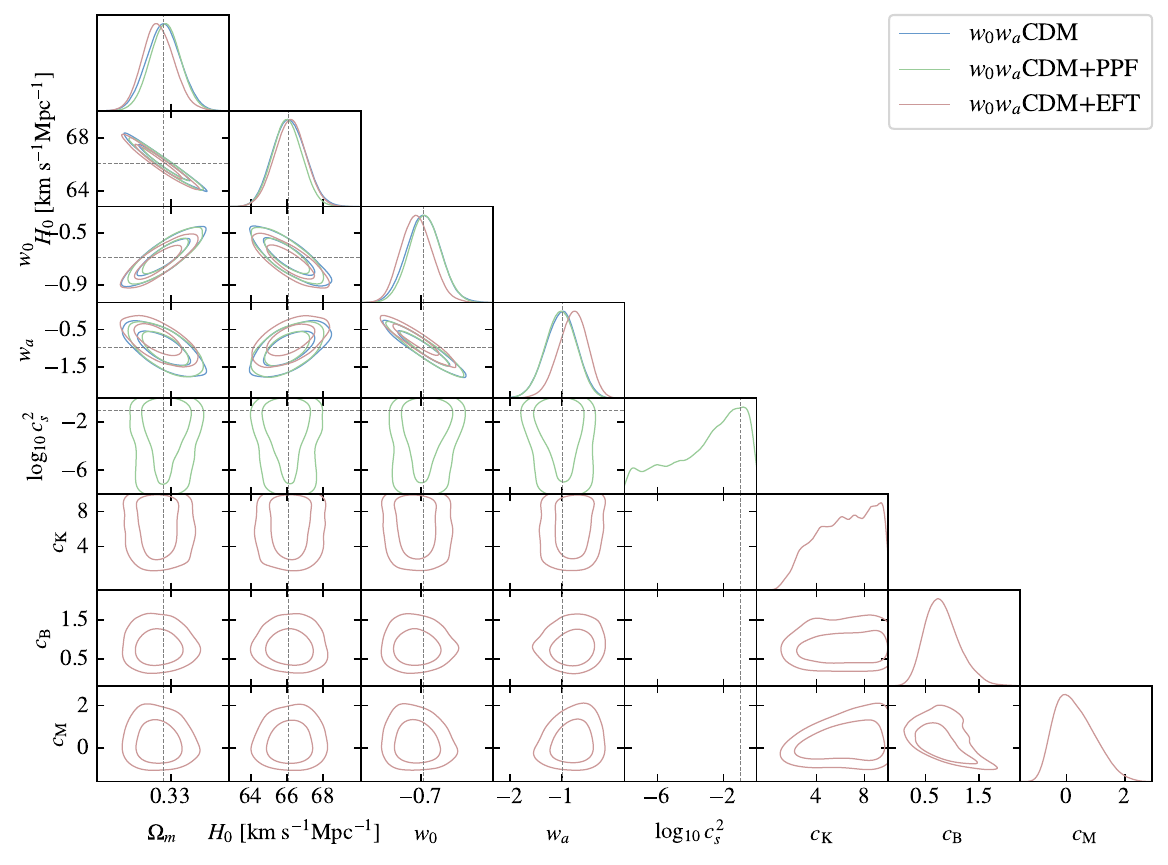}
    \caption{{\it{ Constraints on $w_0w_a$CDM model    
    under  Parameterized Post-Friedmann (PPF) and Effective Field Theory (EFT) 
perturbation descriptions, using BAO+CMB+SNe datasets. The contours represent 
the 68\% and 95\% credible intervals. The results are summarized in 
Table~\ref{table:mcmc_result}. Additionally, the dashed lines represent the 
Maximum A Posteriori (MAP) values for $w_0w_a$CDM+PPF, namely 
    $\Omega_\mathrm{m}=0.3258$, $H_0=66.09$ $\mathrm{km}$ $\mathrm{s}^{-1}$ 
$\mathrm{Mpc}^{-1}$, $w_0=-0.6871$, $w_a=-0.9867$, $\log_{10}{c_s^2}=-0.9780$, 
marginalized   using \texttt{procoli}.}}}
    \label{fig:mcmc_all_w0wa}
\end{figure}

\begin{table*}[ht]
\centering
\caption{Summary  of  cosmological parameter constraints from different dataset 
combinations, for $\Lambda$CDM,  $w$CDM model  and $w_0w_a$CDM models,  under  
PPF and EFT 
perturbation 
descriptions. In each case we show the marginalized posterior means and the 
$68\%$ credible intervals.}
\begin{tabular}{lcccccccc}
\hline
\multirow{2}{*}{Model} & \multirow{2}{*}{$\Omega_\mathrm{m}$} & 
\multirow{1}{*}{$H_0$} & \multirow{2}{*}{$w_0$ or $w$} & \multirow{2}{*}{$w_a$} 
& \multirow{2}{*}{$\log_{10}{c_s^2}$} & \multirow{2}{*}{$c_\mathrm{K}$} & 
\multirow{2}{*}{$c_\mathrm{B}$} & \multirow{2}{*}{$c_\mathrm{M}$}\\
 &  & ($\mathrm{km}\,\mathrm{s}^{-1}\,\mathrm{Mpc}^{-1}$) &  &  & & & & \\
\hline
$\Lambda$CDM & $0.3020^{+0.0038}_{-0.0038}$ & $68.33^{+0.30}_{-0.30}$ & --- & 
--- & ---  & ---& --- & ---\\

$w$CDM & $0.3042^{+0.0059}_{-0.0059}$ & $68.06^{+0.68}_{-0.68}$ & 
$-0.989^{+0.028}_{-0.028}$ & --- & --- & --- & --- & --- \\

$w$CDM+PPF & $0.3052^{+0.0059}_{-0.0059}$ & $67.90^{+0.69}_{-0.69}$ & 
$-0.981^{+0.028}_{-0.028}$ & --- & $-4.3^{+2.3}_{-2.3}$ & --- & --- & ---\\

$w$CDM+EFT & $0.3055^{+0.0060}_{-0.0060}$ & $67.65^{+0.72}_{-0.72}$ & 
$-0.959^{+0.029}_{-0.029}$ & --- & --- & $6.7^{+3.0}_{-1.2}$ & 
$1.12^{+0.33}_{-0.33}$ & $0.25^{+0.55}_{-0.86}$ \\

$w_0w_a$CDM & $0.3258^{+0.0093}_{-0.0093}$ & $66.12^{+0.90}_{-0.90}$ & 
$-0.685^{+0.093}_{-0.093}$ & $-1.01^{+0.30}_{-0.30}$ & --- & --- & --- &--- \\

$w_0w_a$CDM+PPF & $0.3264^{+0.0091}_{-0.0091}$ & $66.02^{+0.89}_{-0.89}$ & 
$-0.683^{+0.094}_{-0.094}$ & $-0.996^{+0.31}_{-0.31}$ & $-3.00^{+2.9}_{-0.99}$ 
& 
--- & --- & ---\\

$w_0w_a$CDM+EFT & $0.3229^{+0.0087}_{-0.0087}$ & $66.18^{+0.86}_{-0.86}$ & 
$-0.724^{+0.081}_{-0.092}$ & $-0.79^{+0.30}_{-0.25}$ & --- & 
$6.4^{+3.4}_{-2.1}$ 
& $0.82^{+0.25}_{-0.37}$ & $0.30^{+0.51}_{-0.83}$\\
\hline
\end{tabular}
\label{table:mcmc_result}
\end{table*}

\end{widetext}

\bibliographystyle{apsrev4-1}
\bibliography{ref}

\end{document}